# Filaments in the TGBA phase

Lubor Lejček*, Vladimíra Novotná, and Milada Glogarová

Institute of Physics, the Czech Academy of Sciences, Na Slovance 2, 182 21 Prague 8, Czech Republic

Abstract

A model of filaments of the TGBA phase arising from the homeotropic smectic-A phase and nucleating on the sample surface is proposed. The model is based on the concept of finite blocks of parallel smectic layers forming a helical structure. The blocks are surrounded by dislocation loops. The model describes the filament structure near the sample surface and describes the observed inclination of the filament axis with respect to the easy direction of the molecular anchoring on the surface. The model is based on the observations of filament textures of the TGBA phase in a new chiral liquid crystalline compound, but can be applied for forming of TGBA filaments in any compound. The compression modulus of the compound has been estimated using such parameters as anchoring energy, estimated from the field necessary to transform the structure into the homeotropic smectic-A, and the observed filament width.

## 1. Introduction

Twist grain boundary (TGB) phases in liquid crystalline compounds with chiral molecules are the frustrated phases existing due to the competing intermolecular interactions and strong molecular chirality, which lead both to the assembly of molecules in layers and the formation of spiral structures. The structures of these phases are inevitably accompanied with defects having the significant consequences on their nucleation, textures and then on theirs properties.

The TGBA phase, which is the object of this study, is composed of blocks (slabs) of the orthogonal smectic-A. Due to chirality they rotate about an axis lying in the smectic layers, the pitch of rotation being typically in the range of the visible light wave length. The blocks (slabs) are separated by systems of screw dislocations forming *twist boundary* analogously as in solids where crystal grains are separated by screw dislocations [1]. The existence of TGB phases was first predicted by de Gennes [2], theoretically described by Renn et al. [3] and afterwards discovered by Goodby [4]. So far there is a lot of papers reporting the existence of TGB phases in various compounds and describing their properties. Typically the TGB phases occur below the Blue phase or the cholesteric phase, but may appear directly below the isotropic phase. There is also a case when TGBA phase appears as a reentrant phase below the smectic-A phase [5].

The textures of TGB phases exhibit various features depending on the sample thickness, surface [6] and geometries [7]. Besides, paramorphic textures from the neighboring phase survive in the TGB phase and in the opposite, the features of TGB textures persist in the non-frustrated phase. The reason is the inevitable presence of defects in the TGB phases, which have to be melted in the other next phases. Generally, the textures of the TGB phases are diverse and complicated. In free standing films, as well as in samples with homeotropic anchoring, the filament [6] or fingerprint textures typically occur [5-12]. Nevertheless, under the homeotropic anchoring also features similar to the fan-shaped textures known from classical smectic phases may be observed in the TGB phases [6,8]. Under the planar anchoring a blurred fan-shaped texture [6], oily streaks structure or irregular grains can appear. In the grain texture the color corresponds to the pitch length of the TGB block rotating, the rotation axis being perpendicular to the sample plane [5-8].

*Corresponding author: lejcekl@fzu.cz

Here we report a model of filaments nucleating from the homeotropic smectic-A phase. We call "filament" a linear object observed e.g. in [4-6]. The model of filament is based on the experiment performed on PHB(S) compound, in which the TGBA phase is the only mesophase and the smectic-A phase is induced by an electric field. The article is structured as follows. Section 2 presents the observed textures of the TGBA filaments in the studied compound. In sections 3 and 4 the geometry of the TGBA filament is described and the energy of the filament is estimated based on the smectic A elasticity. Together with the observed values of filament dimensions this energy permits discussion of the layer compressibility and estimation of the layer compression modulus $B$. Finally, the filament model is used in section 5 to explain the observed orientation of the filament.

## 2. Texture studies

The TGBA textures are studied in different types of the glass cells. Herein, the effect of filaments nucleation is presented for a commercial cell, 5 μm thick, provided with transparent ITO electrodes and with planar anchoring ensured by rubbing a surfactant, with easy direction along the electrode edge. Samples are filled with a new liquid crystalline compound denoted PHB(S) [13] having the enantiotropic TGBA phase in the temperature range from 27ºC to 33 ºC. A typical texture in a planar sample observed in crossed polarizers on the sample cooled down from the isotropic phase is shown in Fig. 1. This texture contains colored grains, the color corresponding to the pitch length of the TGB helix with the helical axis perpendicular to the sample plane. The spectrometric measurements show the pitch length changing from 380 nm (blue color) to 750 nm (red color) on decreasing temperature within the interval from 32.9 ºC to 31.8 ºC; for lower temperature the pitch length is out of spectral range. An electric field of about 60 V per the sample thickness changes this texture to a homogeneously dark state, showing a uniaxial structure with the optical axis perpendicular to the sample surface. This structure corresponds to a well aligned homeotropic smectic A phase. After the field is switched off a filament texture gradually appears (Figures 2a-c). In Figure 2 the edge of the photographs corresponds to the edge of the electrode, being parallel to the orientation of the crossed polarizers. From these figures one can see that the axes of filaments make a certain angle between 45 and 50 deg with the easy direction at the sample surface (electrode edge). It can be concluded that under the field the director becomes oriented perpendicularly to the sample surface (dark state) and then slowly relaxes to a structure enforced by the planar surface anchoring. The filaments thus originate from the homeotropic state established by the field. Let us point out that the filament texture is typical for samples with the homeotropic anchoring or the free standing films [5-12].

Based on the dielectric spectroscopy measurements [13], the permittivity, $\varepsilon$, has been established in a broad frequency range. We have found that at lower frequencies $\varepsilon_\parallel$ is higher than the permittivity measured on the sample without field application, where the component along the short molecular axis $\varepsilon\perp$ prevails. Positive dielectric anisotropy, $\varepsilon_\parallel > \varepsilon\perp$, is the reason for preference of homeotropic alignment under the applied electric field. Thus the transformation of the TGBA phase in the sample with the planar anchoring can be understood as an analogy of the Frederiks transition in nematics [13].

## 3. Filament of the TGBA phase composed of finite smectic blocks

Rotation of blocks forming the TGBA phase, namely the fact they are finite in all dimensions, is basically responsible for the formation of TGB filaments. The blocks with the structure of the smectic A phase rotate along the axis parallel to the smectic layers, the next blocks being relatively rotated by an angle $\omega = 2\pi l_B / P$, where $P$ is the helical pitch and $l_B$ is the dimension of the blocks along the rotation axis. Then the ratio $P/l_B = N$ is the number of blocks over the pitch length, $P$. To calculate the free energy of the TGBA phase let us choose the coordinate system connected with unperturbed smectic-A layers with the $z$-axis oriented along the layer normal, and $x$ and $y$ axes in the plane of the smectic layers. The chiral term responsible for layer rotation was proposed in [14] as $-D_2 \frac{\partial \Omega}{\partial x}$ with $D_2$ as a chiral constant, where $\Omega(x)$ describes the rotation of the blocks around the $x$-axis at the position $x/l_B$ with respect to the system of unperturbed parallel smectic-A layers. This chiral term can be associated with the elastic curvature term of the type $\frac{K}{2}\left(\frac{\partial \Omega}{\partial x}\right)^2$ where $K$ is the curvature elastic constant of the smectic A, which is isotropic in the $xy$-plane. Then the curvature part of the free energy density together with the chiral term can be rewritten as $\frac{K}{2}\left(\frac{\partial \Omega}{\partial x} - q_o\right)^2$ with $q_o = \frac{D_2}{K}$, $q_o$ being connected with the pitch $P$ as $q_o = \frac{2\pi}{P} = \frac{\omega}{l_B}$. The chirality of liquid crystal, i.e. the sign of $q_o$, determines the sign of $\omega$.

In infinite blocks the smectic layers are not deformed. Then the minimum of the free energy density leads to the equation $\frac{\partial \Omega}{\partial x} = q_o$ giving continuous rotation of the smectic layers in the form $\Omega(x) = -\frac{\omega}{2} + \frac{\omega}{l_B}x$. However, the TGB structure consists of discrete rotation of blocks, therefore the form $\Omega(x)$ can be used for geometrical description of TGB phase only when $x = kl_B$, $k \in (0, N)$ being an integer and $N$ describes the number of blocks within the pitch length, i.e. $N = P/l_B$. The angle $\Omega(x)$ is connected with the density of screw dislocations in TGB walls between blocks.

Above we have outlined a well-known geometrical description of the TGB phase [14 - 16]. In the following, we will describe a nucleation of the TGB phase and its geometry in the smectic A liquid crystal with the smectic layers parallel to the sample surfaces. Such a situation has been observed in the experiment described in section 2, when the TGBA phase arises from the smectic A phase induced by the electric field (see Fig. 2). In that case the TGBA phase appears in a form of needles (filaments) which are finite in all three dimensions. In the nucleation of filaments both effects of chirality and the surface anchoring preferring the planar molecular orientation are combined.

A nucleus of a block of TGBA phase emerged inside the smectic A phase at the sample surface is schematically shown in Fig. 3. The smectic layers in the block are inclined by an angle $\Omega$ with respect to the layers in the surrounding smectic A phase. We suppose that layers in a block are straight except for the sample surface, where they are warped creating a surface *wall of the flexion* there [17] with the edge dislocations accommodating the layer curvature near the surface. The layers parallel to the surface terminate near the inclined block by another system of edge dislocations separated by the distance $b/\tan\Omega$ along the $y$-axis, $b$ being the layer spacing. In fact the system of edge dislocations forms a boundary, which can be called an *incoherent twin wall* [18, 19] in analogy with twins in solid crystals. For blocks finite along the $x$-axis the inclination of layers is accommodated by a *twist grain boundary* formed by screw dislocations perpendicular to layers in the block and connecting the edge dislocations with the sample surface (dot-and-dash lines in Fig. 3). Blocks are thus surrounded by parts of dislocation loops starting and ending on the sample surface. The number of the dislocation loops surrounding the block is simply $h/b$ or it can be related to the angle $\Omega(x)$ by the ratio $l/(b/\tan\Omega)$ [18 - 20], $h$ and $l$ being the block height and width, respectively (Fig. 3). On the surface the blocks are ended by a *border line* (Fig. 3). The *border line* also terminates the edge dislocation wall on the sample surface. Note that on the other side, at the distance $l$ from the *border line* (Fig. 3), the block is connected to the system of parallel smectic layers continuously, just by tilt deformation. Such tilt deformation wall is analogous to the *coherent twin wall* [18, 19] in solid crystals. Thus the opposite sides of bocks differ from each other.

The chirality together with the surface anchoring induces a creation of blocks near the surface. The layers in the neighboring blocks are relatively turned by an angle $\omega$. Possible situations are shown in Figs. 4, 5, 6. Fig. 4a shows a case when the neighboring blocks are inclined with respect to the unperturbed layers parallel to the sample surface by angles $-\omega/2$ (dashed lines) and $\omega/2$ (full lines), respectively (this situation arises near $x \approx 0$ or $x \approx P/2$). The block depicted by dashed lines is situated behind the block depicted by full lines. Each block is surrounded by a system of dislocation loops as depicted in Fig.3. For simplicity the dislocations are not shown in the figure. Near the sample surface the layers are curved due to the planar surface anchoring. Schematic drawing of the situation shown in Fig. 4a is presented in a simplified way in Fig. 4b. In Fig. 4b neighboring blocks are shown in perspective represented by just one smectic layer enveloping a block. Layer deformation near surface is neglected.

Fig. 5a shows two neighboring slabs with $\omega/2<\Omega<(\pi-\omega)/2$ i.e. somewhere in the interval $0<x< P/4$. Due to the relative inclination of smectic layers the relative heights of slabs, $h_f$ and $h_b$ could be different. The relative inclination of smectic layers in neighboring slabs can be seen as relative turning of dashed and full lines where the blocks overlap. Again the block depicted by full lines is situated in the front of the block depicted by dashed lines. In perspective, those two blocks are schematically drawn in Fig. 5b. Due to the block rotation the *border line*s of neighboring blocks on the sample surface are displaced by $u$.

The sign of displacement $u$ is determined by the chirality of liquid crystal, i.e. by the sign of angle $\omega$, which can be seen in Fig. 5a. Rotating of neighboring blocks in Fig. 5a is right-handed. The relative displacement $u$ of the *border line* (i.e. displacement of layers depicted by dashed lines from layers represented by full lines on the surface) is oriented in the left direction (see also Fig. 5b). By changing rotation to left-handed the relative displacement $u$ of *border line* will be oriented to the right.

Fig. 6a represents the neighboring blocks inclined with respect to the layers parallel to the sample surface by angles $\Omega = -\frac{\omega}{2}+\frac{\pi}{2}$ (full lines) and $\Omega = \frac{\omega}{2}+\frac{\pi}{2}$ (dashed lines), respectively. Block drawn in dashed lines is situated behind the block in full lines as shown in perspective in Fig. 6b. Such a block arrangement occurs near $x \approx P/4$.

Arrangement presented in Figs. 4, 5, and 6 in a row gives a structure of a TGBA filament. This filament structure of the length $P/2$ is in a simplified 3D view depicted in Fig. 7. In Fig. 7 only the smectic layers enveloping blocks terminating on the sample surface (ends of smectic layers of the width $l_B$ which form a *border line*). Figure 4 represents the situation at the front side of Fig. 7. Fig. 5 shows blocks between the front side and the center of Fig.7. Finally, Fig. 6 corresponds to the situation in the center of Fig. 7, where neighboring blocks are continuously reconnected with smectic layers in opposite directions. Smectic molecules in inclined layers of blocks are also inclined from the normal of the sample surface. Because of the inclination of molecules in filament blocks the filament shows an optical contrast in the polarized light with respect to its surroundings. Note that in Fig. 7 also curvature deformation of layers due to the surface anchoring is shown for illustration.

The local rotation axis between neighboring blocks is parallel to the *x*-axis but its position in *z*-direction above the sample surface can differ from block to block depending on the displacement $u$. Let the position $z_a$ be the position of the rotation axis of layers enveloping neighbor blocks (Fig. 8). In the simplified case of non-deformed smectic layers the position $z_a$ of block rotation axis in the interval $x \in (0, P/4)$ can be determined as $z_a = u \sin\Omega(x) \sin(\Omega(x)+\omega)/\sin\omega$ (see Fig. 8). In the other parts of the interval $(0,P)$ $z_a$ can be expressed in a similar way. The mean value of $\bar{z}_a$ over the interval $(0,P)$ is $\bar{z}_a = \frac{u}{2}\cot\omega$ supposing that *u* does not depend on the x-coordinate. Note that the last relation is based on a very simplified geometrical description of neighbor blocks.

## 4. Energy of TGBA filament

In this section we propose an approximate energy of the TGBA filament over the period *P*. It will enable us to estimate the filament dimensions in directions perpendicular to the rotation axis *x*, namely the height of *k*-th block in the *z*-direction, $h_k$, and its width, along the *y*-direction, $l_k$. Generally, the elastic energies of both screw and edge dislocation walls surrounding each block contribute to the full energy. As for the screw dislocation in the smectic A, recently more detailed solution for a screw dislocation was proposed and a non-zero elastic self-energy was determined [21]. In the limit of an infinite medium, the elastic self-energy given in [21] leads to the Kléman´s term $\frac{Bb^4}{128\pi^3 r_o^2}$ [22], where *B* is the layer compression modulus [14] and $r_o$ is the dislocation core radius. The elastic energy of an edge dislocation per unit length can be written as [17]:

$$E_e = \frac{Kb^2}{2\lambda r_o}, \qquad (1)$$

where $\lambda = \sqrt{K/B}$ with the mean Frank elastic constant *K*. When comparing (1) with the Kléman´s contribution to the screw dislocation energy, it can be found that it is of the order $10^{-3}$ smaller compared with the energy of an edge dislocation and thus can be neglected and only the elastic energy of an edge dislocation walls will be considered.

The length of the edge dislocations in a wall equals to the thickness $l_b$ of a block along the x-direction is $l_B n_k$ where $n_k$ is the number of dislocations in the edge dislocation wall. Assuming smectic layers in blocks to be just inclined with respect to the non-deformed layers parallel to the sample surfaces the edge dislocations in a wall will be uniformly distributed as seen in Fig. 3. The number of edge dislocations is $n_k = l_k / d_k$ and the distance between the edge dislocations along the *y*-direction is $d_k = b/\tan\Omega(k)$ with $\Omega(k) = -(\omega/2) + \omega k$ (Fig. 3 with parameters *l* and *d* changed to $l_k$ and $d_k$, respectively). Then the edge dislocation wall energy can be approximated as:

$$\sum_{k=1}^{N} l_B E_e n_k, \qquad (2)$$

where $P/l_B = N$. The chiral term can be written as $-K\frac{\partial \Omega}{\partial x}q_o$. Taking $\frac{\partial \Omega}{\partial x} = q_o$ this chiral term leads to a decrease of the energy (per unit volume) $-Kq_o^2$ against to the parallel unwound layers. The energy of the chiral term over the period *P* is

$$-Kq_o^2 \sum_{k=1}^{N} l_k h_k l_B, \qquad (3)$$

where $l_k h_h l_B$ characterizes a block volume.

After the electric field is switched off, the filament is nucleated in the vicinity of the sample surface, where the anchoring affects. At the surface molecules are preferentially oriented along the surface easy direction. Being the axis of the chiral rotation of molecules (i.e. the rotation of layers in the smectic A) along the *x*-axis, the orientation of the easy direction on the sample surface will be along *y*-axis. The anchoring energy $W_A$ lowers the total energy by

$$-W_A \sum_{k=1}^{N} l_k \, l_B . \qquad (4)$$

Then the total energy $U$ of the filament is the sum of expressions (2) – (4). In our model the energy $U$ is taken in an isotropic smectic A liquid crystal. It does not take into account the energy of the layer curvature near the sample surface as we suppose the principal curvature energy is concentrated into edge dislocation walls.

The dimensions of a filament $h_k$ and $l_k$ differ with the position of the block along the *x*-axis. In our simplified model we will use mean values $\bar{h}$ and $\bar{l}$ defined as:

$\bar{h} = \frac{1}{N}\sum_{k=1}^{N} h_k$ , $\bar{l} = \frac{1}{N}\sum_{k=1}^{N} l_k$ and $\bar{S} = \frac{1}{N}\sum_{k=1}^{N} l_k h_k$ . Area $\bar{S}$ which is the mean value of the filament cross-sections can be expressed as $\bar{h}\bar{l}/2$, as it can be seen from Fig 3. Therefore, we take $\sum_{k=1}^{N} l_k h_k l_b \approx N l_b \frac{\bar{h}\bar{l}}{2}$, and then

$$U \approx l_B E_e \sum_{k=1}^{N} \frac{h_k}{d_k} - K q_o^2 N \frac{\bar{h}\bar{l}}{2} l_B - W_A N \bar{l}\, l_B . \qquad (5)$$

In the first term of (5) we take the shortest distance between edge dislocations as $d_k \approx b$, which gives the maximum estimation of the edge dislocation wall energy. Taking further $r_o \sim b/2$ [16] and with $N = P/l_B$ we finally obtain the energy of the filament having the pitch length

$$U \approx \frac{KP}{\lambda}\bar{h} - K\frac{2\pi^2}{P}\bar{h}\bar{l} - W_A P \bar{l} . \qquad (6)$$

Moreover, we can suppose that the mean values of the height and the width of the filament are the same, $\bar{h} \sim \bar{l}$.

The nucleation of the filament is supported by thermal activation, the driving force being the influence of both chiral term and surface anchoring. The probability of the nucleation is proportional to $e^{-\Delta U/k_B T}$, where $k_B$ is the Boltzmann constant, *T* the absolute temperature and $\Delta U$ is the energetic barrier of nucleation. The barrier $\Delta U$ is the difference of the maximum energy of filament nucleus with respect to the energy of non-deformed smectic layers having the zero elastic energy. Energy barrier will be estimated using (6). Let us suppose the nucleus in a cubic form with the edge $\sim \bar{h}$. The energy $U_n$ of this nucleus is given by (6) multiplied by ($\bar{h}/P$), which is the factor giving the ratio of nucleus length and pitch, i.e.

$$U_n \approx \left( \frac{KP}{\lambda}\bar{h} - K\frac{2\pi^2}{P}\bar{h}^2 - W_A P\bar{h} \right)\frac{\bar{h}}{P}.$$

The energy barrier $\Delta U$ is obtained for the critical dimension of the filament nucleus $\bar{h}_{cr}$

$$\bar{h}_{cr} = \frac{P^2}{3\pi^2}\left( \frac{1}{\lambda} - \frac{W_A}{K} \right), \tag{7}$$

coming from condition $dU_n/d\bar{h} = 0$. Expression (7) gives the relation between filament dimension $\bar{h}_{cr}$, ratio of the elastic constants $\lambda$ and anchoring energy $W_A$.
Inserting (7) into $U_n$ we obtain

$$\Delta U = U_n(\bar{h} = \bar{h}_{cr}) \approx \frac{KP^4}{27\pi^4\lambda^3}\left( 1 - \frac{W_A \lambda}{K} \right)^3.$$

The anchoring energy $W_A$ can be estimated from the texture observations under the electric field. The homogeneous dark state is reached at about 60 V when the anchored molecules are torn from the surface. The energy of the electric field $E$ in the unit volume of the sample is $-\frac{\varepsilon_o \varepsilon_a}{2}E^2$ where $\varepsilon_a$ is a dielectric anisotropy and $\varepsilon_o$ is the vacuum dielectric constant, $\varepsilon_o = 8.854\times10^{-12}$ F/m. The experiment shows $\varepsilon_a \sim 1$, see Ref. [13] and the thickness of the studied sample is about $t = 5$ μm.

The electric field applied on the sample leads to elastic deformations which become greater towards the sample surfaces because of the surface anchoring. For thin enough samples one can suppose the electric energy is accumulated into the bulk elastic energy which is balanced by the surface anchoring energy $2W_A$ (per unit surface) on both surfaces. For a critical field the elastic energy, which is equivalent to $\frac{\varepsilon_o \varepsilon_a}{2}E^2 t$, the reorientation of molecules at surfaces occurs and

$$W_A \approx \frac{\varepsilon_o \varepsilon_a}{4}E^2 t. \tag{8}$$

From relation (8) we obtain $W_A \sim 1.6\times10^{-3}$ J/m². This energy is comparable with the reported anchoring energy of 5CB nematic liquid crystal on surfaces covered by rubbed polyvinylalcohol film [26].

The pitch $P$ can be estimated as $P \sim 0.5$ μm from the color changes of the texture in Fig.1. With a typical value $K \sim 10^{-11}$ J/m [14] the energy barrier $\Delta U$ is just the function of $\lambda$. When $\Delta U$ is proportional to the energy $k_B T$, the nucleation of filament nucleus starts to be favorable and the parameter $\lambda$ can be estimated. Then the relation $\Delta U \sim k_B T$ with $k_B \sim 1.38\times10^{-23}$ J/K and $T \sim 305$ K gives $\lambda \sim 6.2\times10^{-9}$ m and $B \sim 2.6\times10^5$ J/m³. However, the estimation of $\lambda$ for a typical smectics in [14] is about $\lambda \sim b \sim 3\times10^{-9}$ m, which corresponds to $B \sim 10^6$ J/m³. It means that the studied smectic material PHB(S) is relatively soft, i.e. exhibits higher compressibility. Higher compressibility can be also deduced from comparison of the molecular length and layer spacing [13]. The extended molecule of PHB(S) calculated as 3.1 nm is shorter than the measured layer spacing, which is about 3.6 nm. This difference can be explained by a mutual lengthwise shift of molecules within the smectic layers, which results in higher degree of layer compression in the studied compound [13]. As equation (7) gives the relation among $h_{cr}$, $\lambda$, and $W_A$, the dimension of filament nucleus can be estimated giving $h_{cr} \approx l_B \approx 18$ nm. This value of $l_B$ corresponds to the value given in Ref. [16].

The critical dimension of the filament nucleus $\bar{h}_{cr}$ can be related to the displacement $u$ introduced in the Section 3. Intuitively we can expect that the mean position of the rotation axis of enveloping layers of neighbor blocks is situated nearly in the middle of the mean block height $\bar{h}$. Therefore we take approximately $\bar{h}_{cr} \approx 2\bar{z}_a$ and then $\bar{h}_{cr} \approx u \cot\omega$. This relation shows that both, the critical dimension of filament nucleus and displacement $u$, are related to the energy of dislocation walls, chirality and surface anchoring energy in our model. In the case when the surface anchoring is weak the surface does not influence the filament nucleation. Then the filament nucleation is similar as in the sample bulk and there is no sense to introduce $u$. The filament blocks rotate around the chiral axis identical with the axis of filament.

When the barrier $\Delta U$ is overcome, the parameter $\bar{h}$ has the tendency to increase, $\bar{h} > \bar{h}_{cr}$, and the energy $U$ decreases. Further decrease of $U$ can be also obtained when the nucleus elongates in the direction of the filament axis by a length being multiple of the pitch $P$. Experimental observations show that the elongation of the filament nucleus is much easier along its axis. Our model does not explain this observation but we propose intuitive explication in the following. The increase of block dimension needs the further creation of dislocation loops probably near the surface. Edge segments of the created dislocation loop are generally repulsed from the surface [23] but they should move through the whole block. Edge dislocation segment moving through the filament block is then hindered in its motion by so called Peierls-Nabarro barrier (see e.g. [24]) when crossing smectic layers. (The overcoming the Peierls-Nabarro barrier for dislocations in smectic A by the application of an electric field is discussed in [25]). Moreover, for thin samples the other sample surface also repulses edge dislocations thus preventing block growth. The filament growth in the direction perpendicular to its axis is hindered with respect to the growth along its axis but it is not completely excluded. The filament width $\bar{l}$ can slowly increase so neighbor filaments can touch each other and eventually merge as seen in Figs. 2b, c. Nevertheless, here we do not treat the dynamics of the filament growth, so our comments on filament propagation are just qualitative.

## 5. Orientation of the filament axis

Textures described in section 2 show that the filaments are usually inclined from the easy direction of the anchoring at the sample surface. We want to demonstrate that this inclination is connected with the widths of blocks, $l_B$, and displacements $u$ along the *y*-axis (see Figs. 5, 7). For the purpose of our demonstration we suppose that displacement $u$ of the *border line* exists and the mean characteristic length $L$ of blocks in filament $L = \sqrt{h^2 + l^2}$ (see Fig. 3) is the same for all blocks. The position of the $k$-th block along the $x$-axis can be written as $x_k \approx l_B k$ but we approximate limits of blocks by continuous lines in the $xy$-plane. Blocks in the filament rotate along $x$-axis by an angle $\Omega(x)$. Therefore $l(x) = L\cos(\Omega(x))$ with $\Omega(x) = -\dfrac{\omega}{2} + \dfrac{\omega}{l_B} x$. The projection of the filament blocks on the sample surface is bounded by limits $y_H(x)$ and $y_D(x)$ (Fig. 8). These limits will be determined separately in four intervals of the $P/4$ lengths depending on which side of filament are situated *border lines*. We just remind that blocks terminate at the surface together with edge dislocation wall at the *border line* as seen in Fig.3. The displacement $u$ which is displacement of the positions of *border lines* of neighbor blocks is supposed to be constant. In this model we neglect the layer curvature near the surface shown in Fig. 3. On the other side of *border line* blocks are continuously connected to parallel smectic layers by so called *coherent twin wall*.

Blocks have the width $l_B$ so *border lines* of blocks form stepped line but for simplicity we approximate the *border line* by continuous line $b(x) = \dfrac{u}{l_B} x$. Limits $y_H(x)$ and $y_D(x)$ will be determined in four intervals $(0, P/4)$, $(P/4, P/2)$, $(P/2, 3P/4)$, and $(3P/4, P)$. In these intervals we denote $y_H(x)$ and $y_D(x)$ as $y_{Hi}(x)$ and $y_{Di}(x)$ with subscripts $i$ from 1 to 4. In interval $(0, P/4)$ let us identify the *border line* with $y_{D1}$, i.e.

$$y_{D1}(x) = b(x) + C_{D1}, \tag{9a}$$

where $C_{D1}$ is a constant.

The part of filament in this interval is schematically shown in the lower part of Fig. 7 where blocks are connected with parallel smectic layers on the left side. Then limit $y_{H1}(x)$ is

$$y_{H1}(x) = b(x) + l(x) + C_{H1}, \tag{9b}$$

with $C_{H1}$ constant. Both constants $C_{D1}$ and $C_{H1}$ can be adjusted to zero when the *border line* in (9a) passes through the origin.

In the interval $x \in (P/4, P/2)$ limit $y_{H2}(x)$ is the *border line* because blocks are connected to parallel smectic layers to the right side (see the upper part of Fig. 7) and limit $y_{D2}(x)$ is determined by block projections. Then limits $y_{H2}(x)$ and $y_{D2}(x)$ have forms

$$y_{H2}(x) = b(x) + C_{H2}, \tag{10a}$$

and

$$y_{D2}(x) = b(x) + l(x) + C_{D2}, \tag{10b},$$

with constants $C_{H2}$ and $C_{D2}$. Constants $C_{H2}$ and $C_{D12}$ are determined from the continuity conditions $y_{H1}(P/4) = y_{H2}(P/4)$ and $y_{D1}(P/4) = y_{D2}(P/4)$ at $x = P/4$ Then $C_{H2} = -C_{D2} = L \sin\left(\dfrac{\omega}{2}\right)$. For evaluation of limits at $x = P/4$, we used $P = 2\pi l_B / \omega$. Finally, we can write

$$y_{D2} = \dfrac{u}{l_B} x + L\left(\cos\Omega(x) - \sin\left(\dfrac{\omega}{2}\right)\right) \text{ and } y_{H2} = \dfrac{u}{l_B} x + L \sin\left(\dfrac{\omega}{2}\right). \qquad (11)$$

For $x \in (P/2, 3P/4)$ the *border line* is the limit $y_{D3}$ i.e.

$$y_{D3}(x) = b(x) + C_{D3}, \qquad (12a)$$

and $y_{H3}$ is proposed in the form

$$y_{H3} = b(x) + l(x) + C_{H3}. \qquad (12b)$$

Continuity conditions $y_{H2}(P/2) = y_{H3}(P/2)$ and $y_{D2}(P/2) = y_{D3}(P/2)$ at $x = P/2$ give constants $C_{H3} = -C_{D3} = L\left(\sin\left(\dfrac{\omega}{2}\right) + \cos\left(\dfrac{\omega}{2}\right)\right)$. So we have

$$y_{D3} = \dfrac{u}{l_B} x - L\left(\cos\left(\dfrac{\omega}{2}\right) + \sin\left(\dfrac{\omega}{2}\right)\right) \text{ and } y_{H3} = \dfrac{u}{l_B} x + L\left(\cos\Omega(x) + \cos\left(\dfrac{\omega}{2}\right) + \sin\left(\dfrac{\omega}{2}\right)\right). \qquad (13)$$

Finally, in the interval $x \in (3P/4, P)$ the *border line* is

$$y_{H4}(x) = b(x) + C_{H4}. \qquad (14a)$$

Then limit is $y_{D4}$ is in the form

$$y_{D4}(x) = b(x) + l(x) + C_{D4}. \qquad (14b)$$

Conditions $y_{H3}(3P/4) = y_{H4}(3P/4)$ and $y_{D3}(3P/4) = y_{D4}(3P/4)$ at $x = 3P/4$ give $C_{H4} = -C_{D4} = L\cos\left(\dfrac{\omega}{2}\right)$. Then

$$y_{D4} = \dfrac{u}{l_B} x + L\left(\cos\Omega(x) - \cos\left(\dfrac{\omega}{2}\right)\right) \text{ and } y_{H4} = \dfrac{u}{l_B} x + L\cos\left(\dfrac{\omega}{2}\right). \qquad (15)$$

The filament axis can be described as the mean value of the filament limits $y_D$ and $y_H$, i.e. $y = (y_D + y_H)/2$ in the whole interval $x \in (0, P)$.

In Fig. 9 the projection of filament blocks on the sample plane $xy$ is schematically drawn together with the filament axis, the orientation of which changes along $x$-axis. In our model the filament width, $(y_H - y_D)$, also changes along the axis of block rotation as it is proportional to the projection of $L$ into the plane of the sample surface. The mean angle of filament axis inclination from the easy direction of the planar surface anchoring is determined by the derivative

$$\dfrac{dy}{dx} = \dfrac{u}{l_b} - \dfrac{L}{2}\dfrac{\omega}{l_B}\sin\Omega(x), \qquad (16)$$

in the whole interval $x \in (0, P)$. Then the mean value of the tangent $<dy/dx> = <\tan\alpha>$ over the period $P$ gives $<\tan\alpha> = \frac{1}{P}\int_0^P \frac{dy}{dx}dx = \frac{u}{l_B}$. Therefore the inclination angle $\alpha$ of the filament axis with respect to the $x$-axis is principally given by the ratio $u/l_B$. The inclination from the easy direction is $\frac{\pi}{2} - \alpha$. (see Fig. 9). As the sign of displacement $u$ is determined by the sign of chirality, the angle $\alpha$ has the same property. This is the reason why we observe just one orientation of the filament axis with respect to the easy direction at the surface. The filament nucleated on the other surface have to be inclined from $x$-axis by an angle $-\alpha$. Therefore two sets of filaments nucleating at the both surfaces can be observed (see Fig. 2).

The observations show that the inclination of the filament axis with respect to the easy direction is about 45-50 deg., which gives the displacement $u \approx l_B$. Taking the estimation of the block width $l_B \approx 18$ nm [16], we obtain $u \approx 5b$ with $b \sim 3.6$ nm (see [13]).

In this section we supposed the existence of displacement $u$. The introduction of displacement $u$ is based on geometrical considerations developed in Section 3. The estimation $u \approx l_B \tan\omega \approx 0.23 l_B$ is smaller than $u$ obtained in this Section. The improved estimation of $u$ from the filament nucleation would ask for more detailed model of block shape, including block deformations due to the surface anchoring.

## 6. Discussion

The presented simplified model of TGB filaments is based on the nucleation of the TGB phase on the surface, the liquid crystal chirality and the surface anchoring being effective. Due to a relative rotation of neighbor blocks their ends at the surface are shifted by a displacement $u$. The displacement $u$ determines the position of the rotation axis of neighbor block above the sample surface. We approximately related the displacement $u$ with the critical dimension of the filament nucleus. The existence of $u$ explains why the mean filament axis is inclined with respect to the easy axis of the surface anchoring. Let us point out that such inclination has been observed previously [6], but has remained unexplained. Nucleation can occur on both surfaces so observations show two sets of nearly perpendicular surface filaments (see Fig. 2a).

The anchoring energy can be estimated from this value of the electric field when the transition of the TGBA phase to the homeotropic order with layers parallel to the glass plates is finished, i.e. the sample is completely dark. From the experiments this value has been established to be about 60 V. This value corresponds to the anchoring energy about $1.6 \times 10^{-3}$ J/m$^2$ which is comparable with the measured anchoring energy of 5CB nematic liquid crystal on rubbed polyvinylalcohol film [26].

From equation (7) the compressibility modulus can be determined as $B \sim 2.6 \times 10^5$ J/m$^3$. This value is lower compared to the classical smectic-A phase modulus $B \sim 10^6$ J/m$^3$, showing relatively higher compressibility of the studied liquid crystal in the smectic-A phase. This fact is in accordance with the results obtained from the smectic layer thickness measurements showing that the layer thickness is higher than the length of an extended molecule. Both facts can be explained by a concept of layers composed of molecules associated in pairs, the cores being relatively shifted with respect to each other. Layers with mutually shifted molecules can be compressed more easily than in smectics with the layer thickness comparable to the molecular length.

Principally, filaments can be created also in the sample bulk. In such a case the filament blocks of TGBA phase behave very similarly to bulk twins in solids. The layers in a block are continuously connected to surrounding smectic layers along two opposite faces (coherent twin walls). The other four side faces are enclosed with closed dislocation loops having the edge and screw components as it is schematically depicted in [15]. The screw components of dislocation loops are located between neighboring blocks (twist grain boundaries), while the edge dislocation components discontinuously connect surrounding smectic layers (incoherent twin walls). The axis of the block rotation is identical with the filament axis. Naturally, the creation of a bulk filament is not assisted by the surface anchoring. Such bulk filaments can be nucleated on already grown primary filaments and are probably directed obliquely to the bulk. Due to repulsion of the edge dislocations surrounding the primary and bulk filaments the interaction energy in the mutually perpendicular directions is minimized.

When the anchoring energy is small, the influence of the surface on filament nucleation will be negligible. In such a case the nucleation of the filament on the surface reminds the nucleation of the filament in the bulk.

## 7. Conclusions

We present a model of filaments based on the observations of the chiral liquid crystal PHB(S) exhibiting the isotropic-TGBA phase sequence without any intermediate cholesteric or blue phases. Under an applied electric field the TGBA phase is transformed to the smectic-A phase with the homeotropic structure, which is homogeneously dark in crossed polarizers, showing thus quite perfect alignment. The model describes gradual arising of the TGBA phase in the form of filaments from the field induced homeotropic smectic A phase after the field is switched off. This process starting at the sample surface is driven by chiral forces in combination with the planar anchoring.

Finite blocks of TGBA structure, which compose a filament, are separated from neighboring filament blocks by dislocation loops having screw and edge components and starting and finishing on the sample surface. Screw components of the dislocation loops form twist grain boundaries between blocks in a filament. Edge components of the dislocation loop form incoherent grain boundary between a block and homeotropic smectic-A layers on one side of the block while the other side is continuously connected to surrounding homeotropic smectic A layers (coherent grain boundary) as seen in Fig. 3.

From the value of the surface anchoring, $W_A$, estimated from the experiment, relatively high compressibility of the studied liquid crystalline compound can be deduced. The model of the filament is also able to explain the observed orientation of the filaments with respect to the easy direction of the anchoring on the sample surface. It is worth pointing out that the creation of similar filaments is typically observed in other compounds during the smectic-A - TGBA phase transition under a temperature change and up to now has not been theoretically described and explained.

**Acknowledgements**

This work was supported by the Czech Science Foundation (project 15-02843S).

**Figures and Figure Captions:**

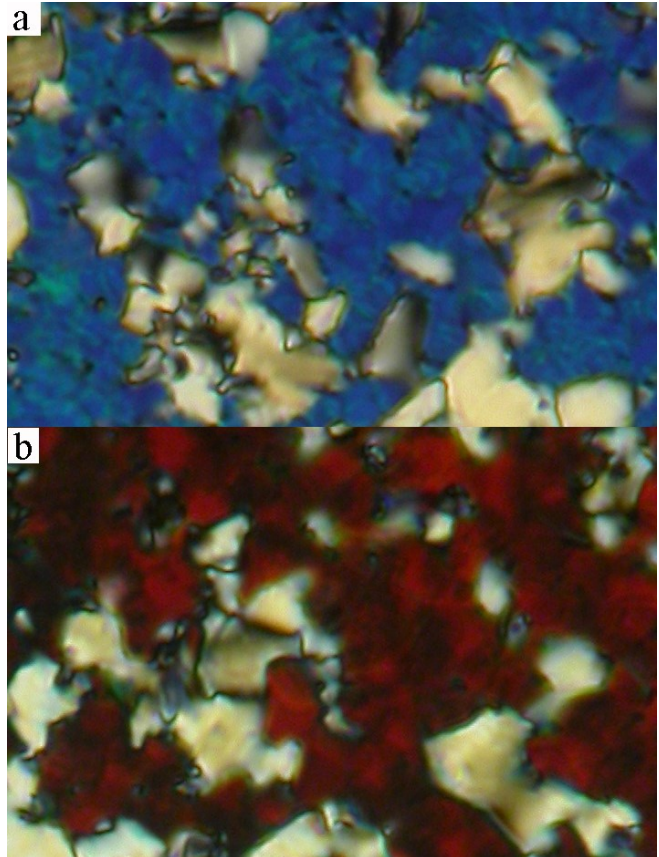

Figure 1: Nucleation of the TGBA phase in PHB(S) compound when cooling down the sample. (a) TGBA phase exhibits blue color at the temperature 33ºC, (b) the color of TGBA phase changes to red one at the temperature 32.2ºC. The width of every micrograph is about 120 μm.

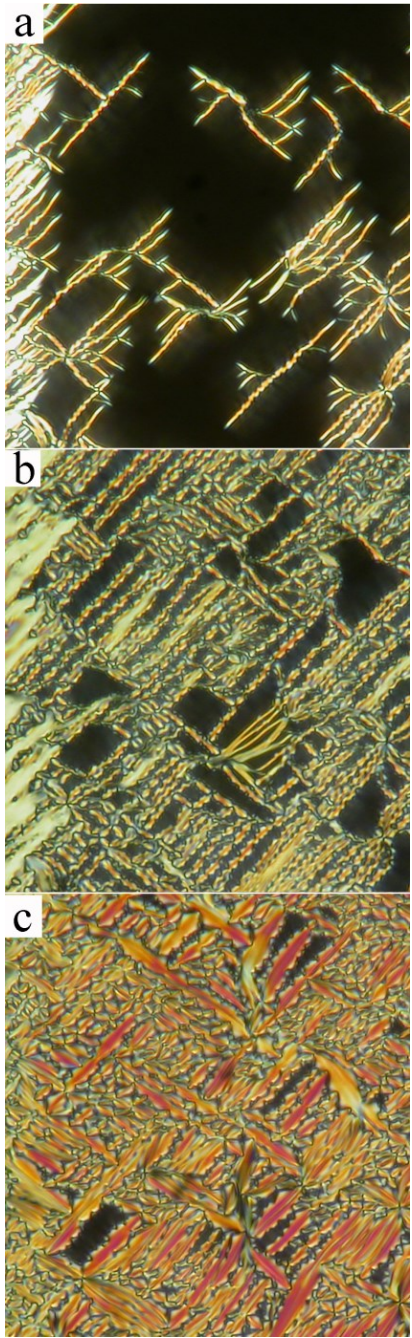

Figure 2: The growth of filaments of the TGBA phase in the homeotropic state of PHB(S) compound after switching off the electric field, temperature 32.2°C.

(a) Individual filaments grow with time in the direction making an angle between 45 and 50 deg. with the easy direction at the sample surface. The easy direction is parallel to the edge of electrode parallel with the edge of the Figure. (b) The density of filaments increases with time, sometimes also filaments perpendicular to the primary system of filaments occur. (c) When filaments cover the whole sample, an individual filament can coalesce making wider areas of TGBA phase. The width of every micrograph is about 150 μm.

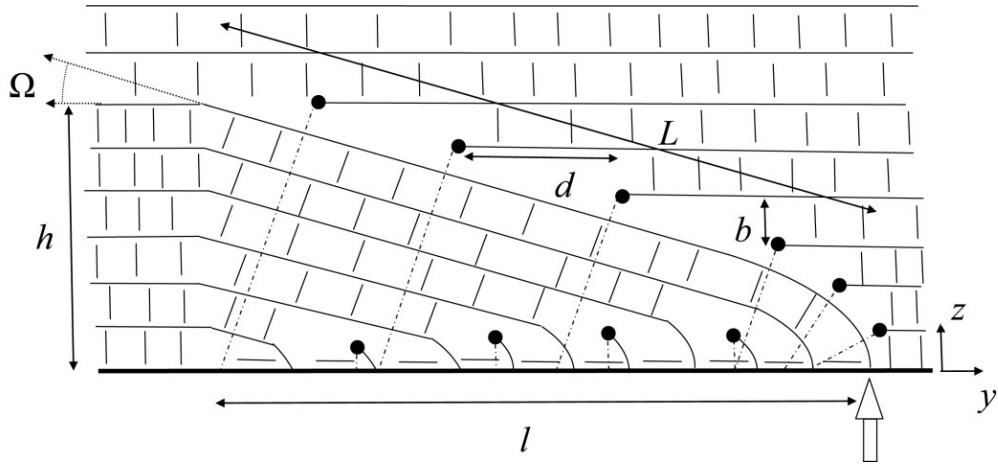

Figure 3: Schematic representation of a block with the smectic A layers inclined by an angle $\Omega$ with respect to layers parallel to the sample surface (thick line). The smectic layers are represented by thin lines with molecules shown as short line segments. The full dots show cross-sections of edge dislocations, which are connected with the sample surface by a system of screw dislocations (dot-and-dash lines). The open arrow shows the position of a *border line* of the inclined block at the surface, the line being parallel to the *x*-axis. Dimensions of block are the height *h* and width *l* determining the characteristic length *L* as $L=\sqrt{h^2+l^2}$. Distance between edge dislocations is denoted as *d* and the smectic layer thickness is *b*. The *y* and *z*-axes are indicated, the *x*-axis is perpendicular to the plane of figure.

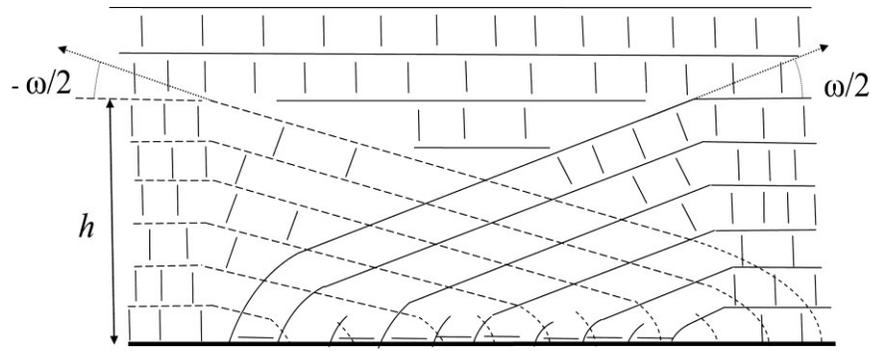

a)

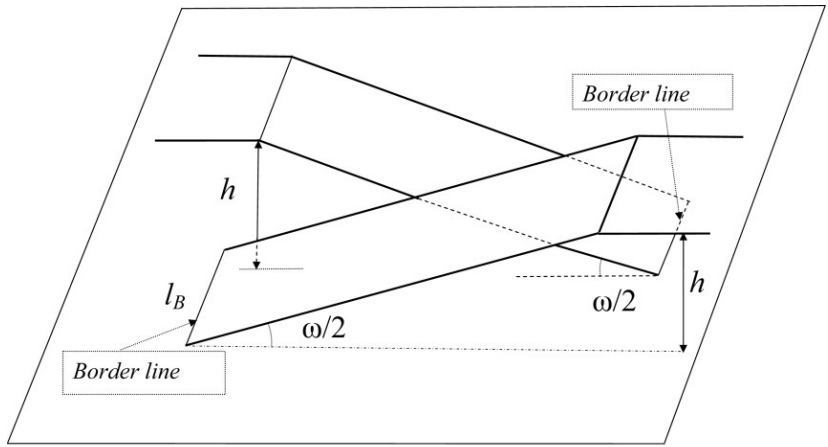

b)

Figure 4: Schematic representation of two neighboring blocks shown (a) in full (block in the front) and dashed lines (block behind) and in perspective (b). The drawing corresponds to the positions at *x~0* and *x~P/2* along the filament.

(a) Blocks are inclined by an angle $\frac{\omega}{2}$ and $-\frac{\omega}{2}$ with respect to the parallel smectic layers and thus mutually inclined by $\omega$. The height, *h*, of both blocks is equal.
(b) Depicted smectic layers constitute an envelope of the neighbor blocks only. For simplicity curvature deformations of layers in blocks near the sample surface are omitted.

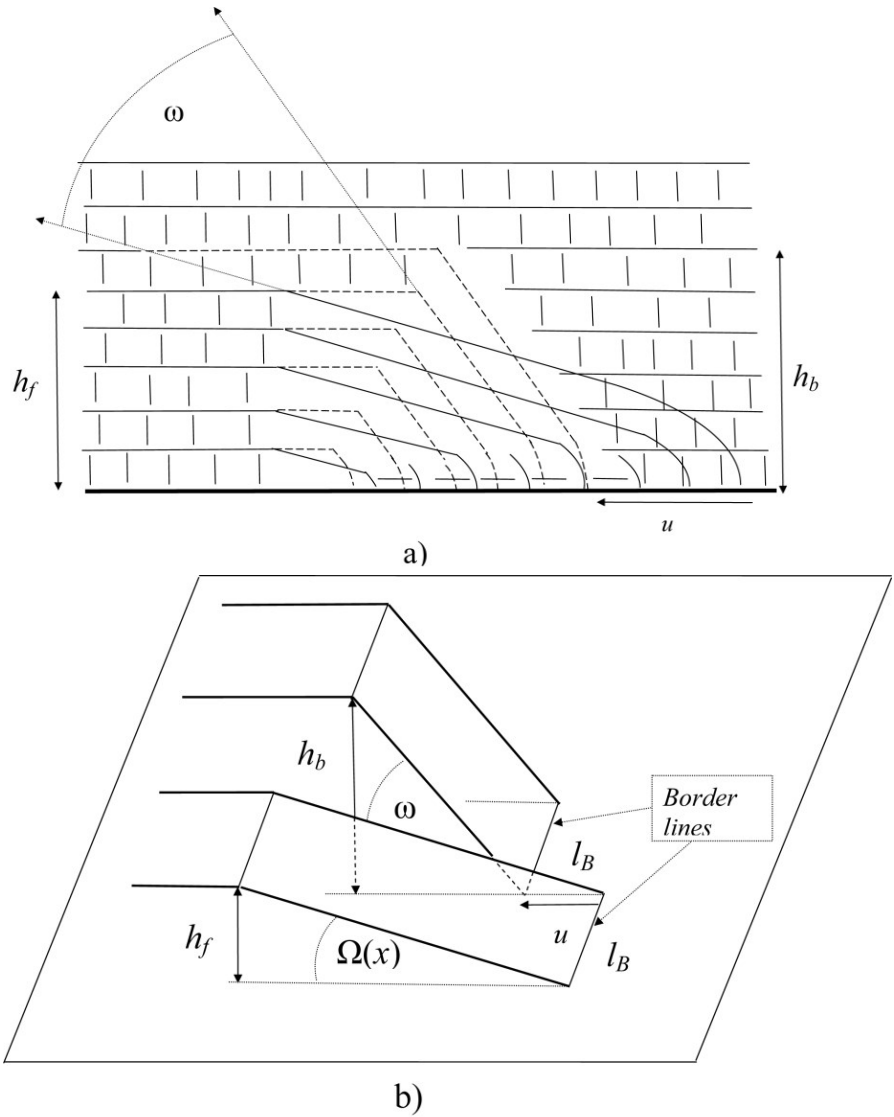

Figure 5: Schematic representation of two neighboring blocks shown (a) in full (in the front) and dashed lines (behind) and in perspective (b). The drawing corresponds to the positions for $x$ between $x=0$ and $x=P/4$. The height of block in front, $h_f$, differs from the height of block behind, $h_b$. The relative displacement of *border lines* of blocks is denoted as $u$.

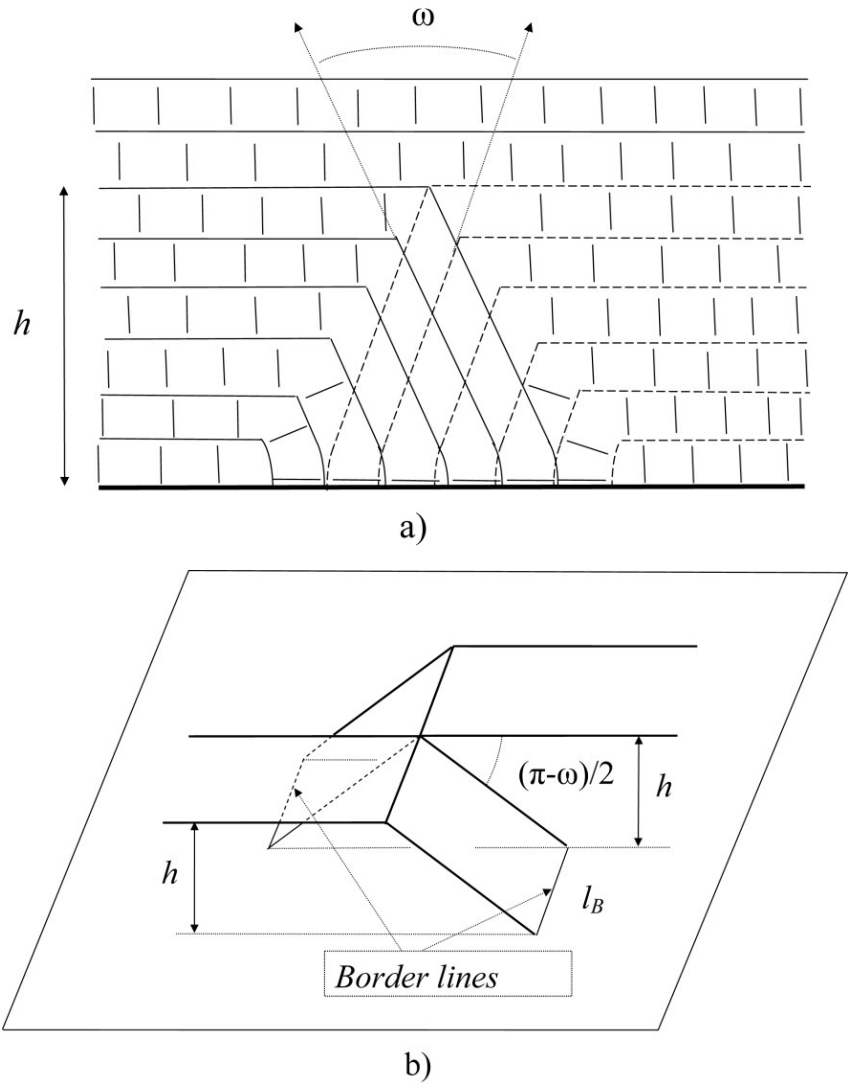

a)

b)

Figure 6: Schematic representation of two neighboring blocks shown (a) in full (in the front) and dashed lines (behind) and in perspective (b). The drawing corresponds to the position at *x=P/4* along the filament. Blocks are inclined by an angle $\frac{\pi - \omega}{2}$ with respect to the parallel smectic layers and thus relatively inclined by $\omega$. The height, *h*, of both blocks is equal.

The same blocks as in (a) are shown in perspective in (b). Depicted smectic layers constitute an envelope of the neighbor blocks only. For simplicity curvature deformations of layers in blocks near the sample surface are omitted.

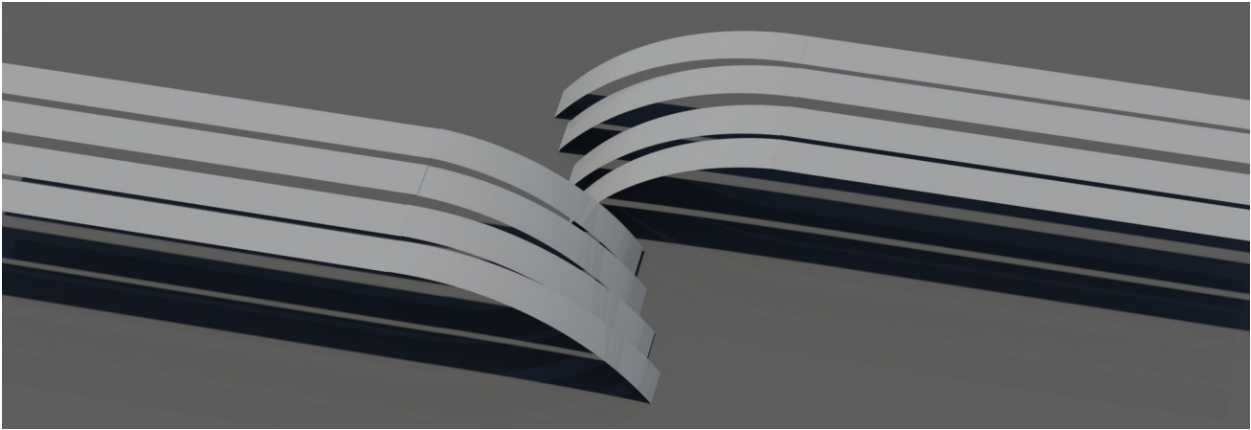

Figure 7: 3D drawing of the smectic layers which terminate on the sample surface tracing there a *border line*. Depicted smectic layers constitute an envelope of the filament only. The presented simplified view demonstrates namely a relative displacement of neighboring blocks on the sample surface caused by chirality induced relative block rotation. The filament segment presented has the length of *P*/2. Note that blocks in the lover part of the Figure, corresponding to the interval $0 \leq x \leq P/4$ are continuously reconnected with parallel smectic layers in the left while blocks in the upper part of the Figure ($P/4 \leq x \leq P/2$) are continuously reconnected with parallel smectic layers in the right.

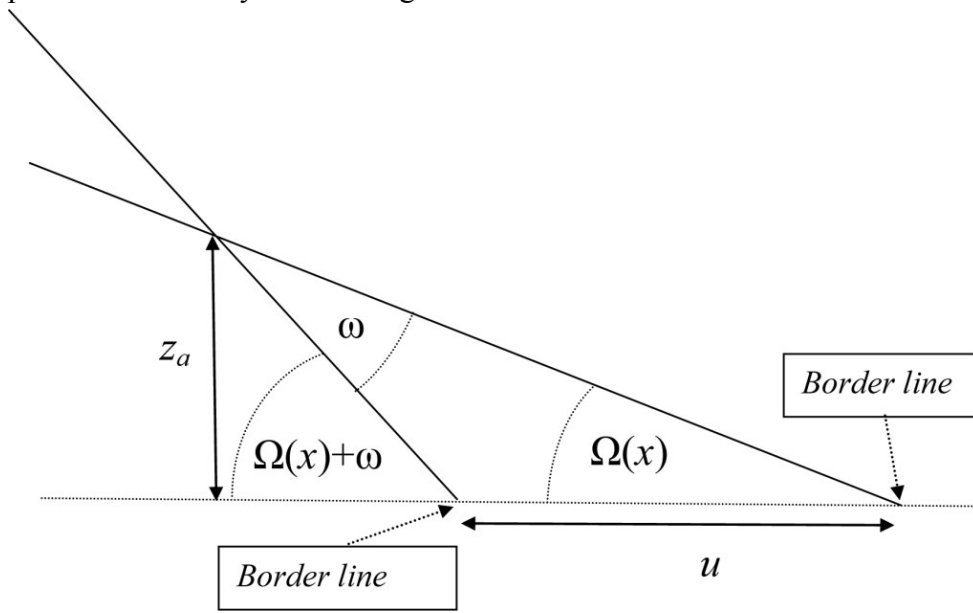

Figure 8: Scheme drawing showing the position of the rotation axis $z_a$ of neighbor filament blocks related to the displacement $u$ of block *border lines*. Enveloping smectic layers are relatively inclined by an angle ω while neighbor block are inclined with respect to the sample surface by angles $\Omega(x)$ and $\Omega(x)+\omega$, respectively. It is a simplified situation not taking into account the layer deformation near the surface due to the anchoring.

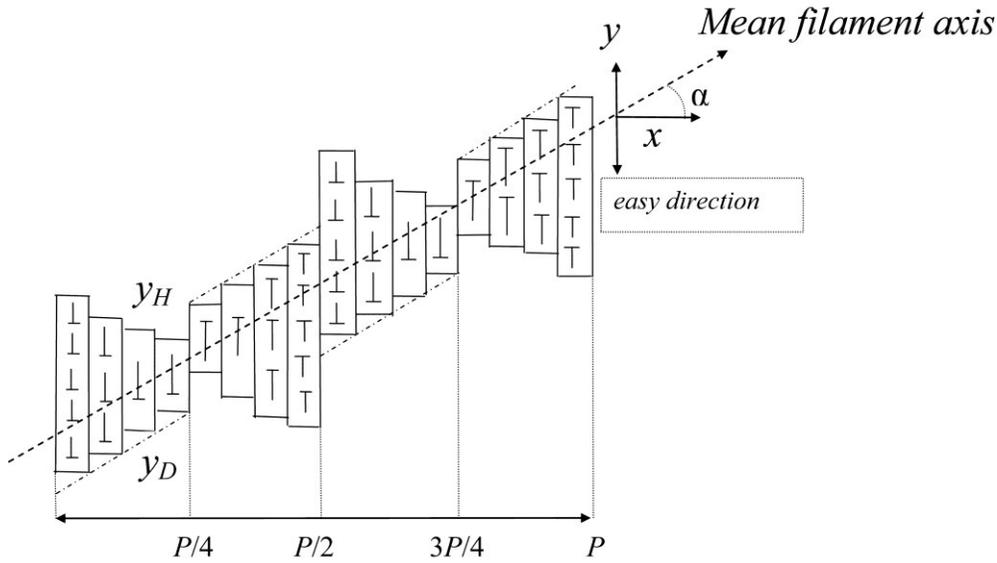

Figure 9: Schematic drawing the filament seen along the normal to the sample surface. The filament segment is shown over the length of pitch *P*. Block rotations lead to the inclination of molecules from the sample surface normal. Inclined molecules are represented by nails the points of which are oriented toward the observer. The length of the nails corresponds to the projection of the inclined molecule to the sample plane. Relative block rotations are illustrated by nails of changing lengths in neighboring blocks. The helix axis is parallel to $x$-axis. Along the helix axis the molecules rotate by $2\pi$. Filament blocks projection onto *xy*-plane is limited by $y_D$ and $y_H$ which define the shape of filament. Mean directions of *border lines* which limit filament segments are denoted by dotted-and-dashed lines. In intervals of the length *P*/4 *border lines* are either $y_D$ or $y_H$. The mean filament axis (dashed line) is inclined with respect to $x$-axis by an inclination angle $\alpha$. The easy direction of the surface anchoring is parallel to *y*-axis.